\documentclass[pra,aps,superscriptaddress,twocolumn,showpacs,10pt]{revtex4-1}
\usepackage[utf8]{inputenc}
\usepackage[T1]{fontenc}
\usepackage{amsmath}
\usepackage{amsfonts}
\usepackage{amssymb}
\usepackage{color}
\usepackage{graphics,graphicx}
\usepackage{epstopdf}

\begin{document}
\title{Optimization of periodic single-photon sources based on combined multiplexing}
\author{Ferenc Bodog}
\affiliation{Institute of Physics, University of P\'ecs, H-7624 P\'ecs, Ifj\'us\'ag \'utja 6, Hungary}
\author{Peter Adam}
\email{adam.peter@wigner.mta.hu}
\affiliation{Institute of Physics, University of P\'ecs, H-7624 P\'ecs, Ifj\'us\'ag \'utja 6, Hungary}
\affiliation{Institute for Solid State Physics and Optics, Wigner Research Centre for Physics, Hungarian Academy of Sciences, H-1525 Budapest, P.O. Box 49, Hungary}
\author{Matyas Mechler}
\affiliation{MTA-PTE High-Field Terahertz Research Group, H-7624 P\'ecs, Ifj\'us\'ag \'utja\ 6, Hungary}
\author{Imre Santa}
\affiliation{Institute of Physics, University of P\'ecs, H-7624 P\'ecs, Ifj\'us\'ag \'utja 6, Hungary}
\author{M\'aty\'as Koniorczyk}
\affiliation{Institute of Mathematics and Informatics, University of P\'ecs, H-7624 P\'ecs, Ifj\'us\'ag \'utja 6, Hungary}
\date{\today}
\begin{abstract}
  We consider periodic single-photon sources with combined
  multiplexing in which the outputs of several time-multiplexed
  sources are spatially multiplexed.  We give a full statistical
  description of such systems in order to optimize them with respect
  to maximal single-photon probability. We carry out the optimization
  for a particular scenario which can be realized in bulk optics and
  its expected performance is potentially the best at the present
  state of the art. We find that combined multiplexing outperforms
  purely spatially or time multiplexed sources for certain
  parameters only, and we characterize these cases. Combined
  multiplexing can have the advantages of possibly using less
  nonlinear sources, achieving higher repetition rates, and the
  potential applicability for continuous pumping. We estimate an
  achievable single-photon probability between 85\%\ and 89\%.
\end{abstract}
\maketitle

\section{Introduction}\label{s:intro}

Applications in quantum information science~\cite{knill, kok, gisin, scarani, duan, sangouard, bennett, bouwmeester, merali, koniorczyk, spring, broome, tillmann} and quantum optics~\cite{cgerry, lund, he, adam, lee} generate an intensive research interest aiming at the construction of periodic single-photon sources (PSPS). Beside the deterministic single-photon sources based on various single quantum emitters such as single atoms \cite{hijlkema, mckeever}, ions \cite{barros, keller}, molecules~\cite{lounis, lettow}, diamond color centers~\cite{beveratos, gaebel, Wu}, and quantum dots~\cite{santori, strauf, polyakov}, probabilistic single-photon sources offer an alternative way to address this problem. This approach is based on the generation of correlated photon pairs. The detection of one of the members of the pair, usually termed as the \textit{idler}, heralds the presence of the other one, referred to as the \textit{signal}. In the literature there are two typical ways of realizing a heralded single photon source (HSPS) based on correlated photon pair generation. The two physical phenomena applied for pair generation  are spontaneous four-wave mixing (SFWM) in optical fibers \cite{Smith2009,Cohen2009,Soller2011} and spontaneous parametric down-conversion (SPDC) in bulk crystals \cite{Mosley2008, Zhong2009, Evans2010, Brida, Broome} or waveguides \cite{Fiorentino2007, Eckstein2011}. These processes can yield highly indistinguishable single photons in an almost ideal single mode with known polarization \cite{Mosley2008,Evans2010, Eckstein2011,Soller2011,Fortsch}.

The major issue of these sources is the probabilistic nature of pair
generation. Though the periodicity can be ensured by periodic pumping, the
number of the generated photon pairs still remains uncertain.  In the
case of pulsed SPDC based HSPS there is a theoretical limit of
single-photon probability $P_1\simeq37\%$ (assuming Poissonian
statistics for the generation of photon pairs), which is insufficient for most of the applications. 

One way to overcome this problem and increase the single-photon
probability is spatial multiplexing in which several HSPSs are used in
parallel \cite{Migdall, Shapiro}. The decrease
%in
of the intensity of each source improves the
single-photon probability compared to that of multi-photon
presence. On the other hand the absence of photons becomes more
likely, too. Multiplexing compensates for this latter by making use of
one of the photons generated in either of the sources.  In principle,
the increase in the number of the sources and the decrease
%in
of their intensity improves the single-photon probability. In an ideal
lossless system this probability tends to one asymptotically. Losses,
however, impose a limitation on this approach. In addition, the growing
number of required HSPSs appears as a drawback in an experimental
implementation. Spatial multiplexing has been realized in experiments
indeed, yet with only up to four heralded single-photon
sources~\cite{Zotter, Collins, Meany,Xiong2015}.

Another possible way of enhancing single-photon probability is time
multiplexing.
Compared to spatial multiplexing, the role of the multiplexed unit is
overtaken by time windows in this case, otherwise the basic idea is
the same. The heralded pulse should leave the time-multiplexed source
precisely at the end of the time period, thus a proper delay should be
introduced. The controlled delay system can be realized with a
storage cavity or loop \cite{Pittman, Jeffrey,FJones2015, Rohde2015, Kaneda2015} or with binary division
strategy \cite{Schmie, Mower, Adam}. Time multiplexed arrangements can be pumped either
with pulses or continuously. The latter may have benefits for
obtaining a real single-mode source of indistinguishable photons. The
increase of the time windows, which is necessary in this system in
order to improve the single-photon probability, however, introduces a
fundamental limitation in the achievable repetition rate.

In actual experimental realizations, the applied optical elements are
not ideal; losses have to be taken into account \cite{Mower,Adam,Bonneau}. In Ref.~\cite{Adam}
we have introduced a theoretical framework describing all the spatial
and time multiplexed single-photon sources realized or proposed thus
far. Our statistical description takes into account all the possible
relevant loss mechanisms. We have shown there that multiplexed sources
can be optimized to reach maximal single-photon probability. This can
be achieved by the appropriate choice of the number of multiplexed
units of spatial multiplexers or multiplexed time intervals, and the
input mean photon number. Furthermore, a novel time-multiplexed scheme based on an SPDC source was proposed by us, which can be realized in bulk optics. This system could
provide a single-photon probability of 85\% with a choice of
experimentally feasible loss parameters.

The ultimate goal of this line of research is to improve the
single-photon probability in realistic systems. Hence, as a logical
continuation of the outlined
%? or antecedents
antecendents, in this paper we consider
\emph{combined} multiplexing: the simultaneous application of both
approaches in the same arrangement. Though the idea of combining
spatial and time multiplexing has already been introduced in the
literature \cite{Glebov, active, Latypov}, a full statistical analysis
of these systems has not yet been performed. In Ref.~\cite{Glebov} the
authors have carried out a Monte-Carlo simulation and optimization of
a combined multiplexing arrangement, in which the outputs of several
storage cavity time multiplexers are spatially multiplexed. In their model,
however, losses of the spatial multiplexers were ignored.
Reference~\cite{active}, presenting actual experiments, includes an
analysis of rather special arrangements, including only a single
SPDC source, but pumped from two sides, which is equivalent to
the application of two independent nonlinear sources.  Reference~\cite{Latypov}
focuses on the study of a time multiplexer using variable optical
delay lines (instead of binary division networks).
The arrangements studied in the latter two papers also contain a special kind of
combined multiplexing in which the output of spatial multiplexers is
multiplexed in time. The possible drawback of such hybrid systems is that they can be pumped with pulses only.

In the present paper we analyze the most general scheme of combined
multiplexing. We assume that the outputs of several time-multiplexed
sources are spatially multiplexed. These kind of combined multiplexers reserve
the advantage of the time multiplexers that they can be pumped continuously. We give a detailed statistical description of combined multiplexing taking into account the possible loss mechanisms. The derived expressions are applicable for combined systems containing any kind of time and spatial multiplexers. Our statistical description can be used for optimizing the setup with respect to single-photon probability. We analyze in detail a particular arrangement which can be realized in bulk optics and performs potentially the best at the present state of the art. We show how combined multiplexing can overcome the issues of the number of required nonlinear photon pair sources in spatial multiplexing, and repetition rate in time multiplexing.
We characterize the cases for which combined multiplexers outperform purely spatially or time multiplexed sources concerning single-photon probability.

The paper is organized as follows. In Sec.~\ref{s:combined} we
describe general combined multiplexing systems, and we also introduce
the particular one which we shall study in more detail.
Section~\ref{s:stat} is devoted to the statistical description of
combined multiplexers, while in Sec.~\ref{s:res} our results regarding
their optimization are described in detail. Finally, in
Sec.~\ref{s:concl} our results are summarized and conclusions are drawn.

\section{Combined multiplexing}\label{s:combined}

\begin{figure}[tb]
\includegraphics[width=\columnwidth]{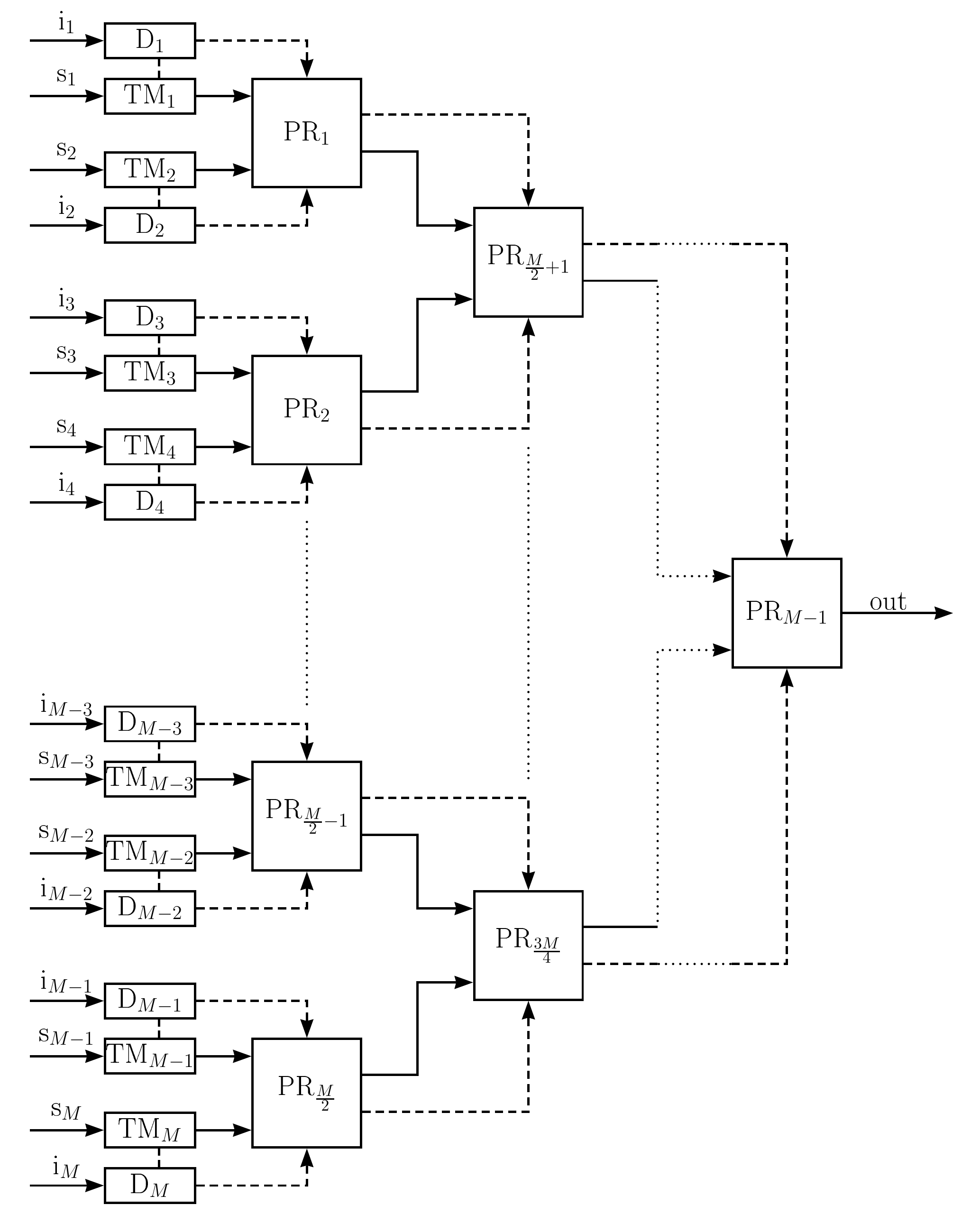}
\caption{\label{f:combinedfig} Scheme of the combined multiplexer. TM$_k$ is the $k$th time multiplexer, D$_k$-s are detectors, PR$_k$-s are photon routers. i$_k$ and s$_k$ denotes the idler and signal arm of the $k$th nonlinear photon pair source. Dashed lines represent electronic control lines.}
%Az kontrolvonalak rosszak. Nem a fotonrouterek irányítják ergymást.
\end{figure}

The idea of combined multiplexing is to use the output of several time
multiplexing arrangements as inputs of a spatial multiplexer in order
to realize a periodic single-photon source. The general scheme is
depicted in Fig.~\ref{f:combinedfig}. In the figure TM$_k$ denotes the $k$th
time multiplexer while the spatial multiplexer is realized by the
photon routers PR$_i$. The arrangement is fed by $M$ nonlinear, e.g.,
SPDC sources, any of them producing completely correlated photon pairs in two modes. In the $k$th arm the idler mode i$_k$ is directed to the detector $D_k$ while the signal mode s$_k$ enters the time multiplexer TM$_k$. The scheme of the time multiplexer is not specified in this configuration, any of the known types can be used. The details of different time multiplexers is described in Refs.\ \cite{Pittman, Jeffrey, FJones2015, Rohde2015, Kaneda2015,Schmie,Mower,Adam}.

The operation of the arrangement can be summarized as follows.
The idler mode i$_k$ is detected by a detector D$_k$ within measurement
time intervals (time windows) of length $\Delta t$. In the case of pulsed pumping $\Delta t$ is equal to the pumping period while for continuous pumping the detector is active for such periods. The observation time covered by $N$ time windows is less  than or equal to the desired period $T$ of the PSPS ($N\Delta t\leq T$).
As the length of the time window $\Delta t$ evidently has a minimal value determined by the characteristics of the system, the number of applied time windows limits the repetition rate of the signal of the time multiplexer.

When the detector fires, the presence of a number of photons is ensured in the given time window in the signal mode s$_k$. These heralded photons enter the time multiplexer that delays them appropriately to arrive at its output at the end of the time period $T$. Hence, the output of a single unit TM$_k$ consists of a periodic train of photons with a period of $T$. The number of photons $i$ is random in each cycle, described by a probability distribution $P(i)$. Usually, the probability $P(0) > 0$ resulting from the case when no photons are detected during the whole period and due to the losses of the multiplexer.
We assume without loss of generality that the time multiplexers in the combined multiplexer are identical and their time period is synchronized.

The outputs of the time multiplexers are directed to the inputs of the spatial multiplexer realized by a sequence of routers.
A photon router PR$_i$ has two input ports and a single output. Combining
multiple routers, a spatial multiplexer with a number of input ports
being powers of 2 and a single output port can be realized, as it is
presented in Fig.\ \ref{f:combinedfig}. We note that there exists a
chained arrangement of sources and routers, without this limitation
of the number of inputs \cite{Bonneau, Mazzarella}. The operation of
the spatial multiplexer is governed by a priority logic which forwards
just a single input mode to the output.
%Nincs rajta az ábrán.

\begin{figure}[b]
\includegraphics[width=\columnwidth]{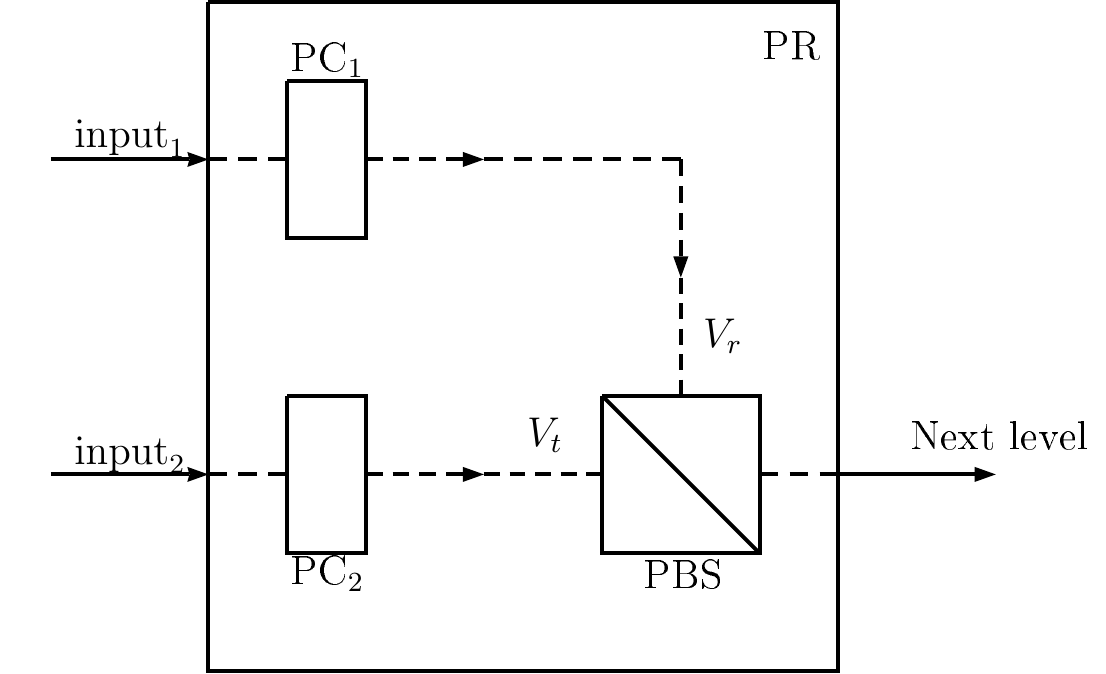}
\caption{\label{f:router} Scheme of the bulk optical photon router. PCs denote Pockels cells, PBS is a polarizing beam splitter. $V_r$ denotes the reflection efficiency and $V_t$ the transmission efficiency of the PBS.}
\end{figure}

From the point of view of our previous results stating that a time
multiplexer built with bulk optical elements can have the highest
single-photon probability, it seems to be interesting to analyze
spatial multiplexing realized with bulk optics as well. 
Accordingly, we will consider a photon router realized in the way depicted in Fig.~\ref{f:router} for the detailed analysis of a particular setup presented in Sec.~\ref{s:res}. This router contains Pockels-cells PC and a polarizing beam splitter PBS.
The chosen mode is selected by the PBS according to the polarization
set by the control logic via the PC-s. As the reflection efficiency $V_r$ and transmission
efficiency $V_t$ of a PBS are generally different, each arm of the  whole spatial multiplexer built from these blocks will have a given, possibly different transmission probability.

At the end of the time period $T$ it is likely that there are more than one time multiplexers from which heralded signal photons are expected to arrive at the corresponding input of the spatial multiplexer. The detectors provide information on the input arm and also the time window in which heralding event occurred. The priority logic of the spatial multiplexer is responsible for forwarding only one of the input modes where the presence of signal photons is predicted by the detector to the output. Taking into account the special characteristics of the spatial multiplexer described above, this control logic has two options for determining the priority. It can simply choose the mode in which a detection event first occurred ignoring the fact that the arms of the spatial multiplexer can have different losses.
It seems obvious, however, that the logic should rather choose the arm of the spatial multiplexer with the highest net transmission probability (i.e., lowest loss).
Our theoretical description presented in the next section shall cover both of these options.

\section{Statistical description of combined multiplexers}\label{s:stat}

In what follows we set up the theoretical framework to calculate the
performance of the combined systems in argument.  First we describe
our improved combined multiplexing system, in which the spatial
multiplexer arm with the lowest loss is chosen.
For a practical realization we may label the time multiplexers
in an order of increasing loss parameters of the corresponding arms of the spatial multiplexer. Thus at
the end of a time period the output of the time multiplexer having lowest labeling number and producing heralded photons expectedly, is directed to
the output. Now, there are two possibilities. If these labeling
numbers correspond to different losses, then the logic will
automatically choose the lowest loss. If the labeling numbers
correspond to the same loss, then the logic simply chooses any of the
multiplexers where heralding event occured, say, e.g. the one with the lowest
label.

Assume that $M$ (power of 2) time multiplexers are spatially multiplexed, and each of
them has $N$ time windows.  For a given time window, let us denote by
$P_0$ the probability of the event that no photon is detected, and
let $P_j$ be the probability of the event that $j$ signal photons
enter the system from the signal mode of the nonlinear photon pair source upon a
detection event in the idler. We calculate the probability $P^{(i)}$
that exactly $i$ photons emanate from the output of the whole
arrangement in a single period. We have, from elementary considerations,
\begin{multline}
\label{eq:ftm0}
P^{(i)}=P_0^{MN}\delta_{i,0}+\\ \sum_{j=1}^{\infty}\sum_{k=1}^{M}\sum_{n=1}^N \binom{j}{i} P_0^{N(k-1)}P_0^{n-1} P_j V_{nk}^i(1-V_{nk})^{j-i}.
\end{multline}
The first term contributes only to the probability $P^{(0)}$
corresponding to the case where no photons are detected during the
whole period. The second term stands for the case when, even though
there are $j$ photons emerging from the nonlinear source of the $k$th time
multiplexing unit in the $n$th time window, only $i$ of them reaches
the output due to the losses of the multiplexing system. The powers of $P_0$ 
correspond to the choice of the priority logic under consideration, that is, 
accepting the signal photon emerging from the time multiplexer with the lowest 
labeling number $k$. The $N(k-1)$st power of $P_0$ describes the case when no
photon pairs were produced in $k-1$ sources in the whole time period while the
$(n-1)$st power means that the heralded photons appeared in the $n$th time
window of the $k$th source. The summations
go over all the possible values of the number of incoming heralded photons $j$,
spatially multiplexed time multiplexers $k$ and time windows $n$. Losses are
described by the parameters $V_{nk}$: the net transmission (i.e.,
total probability of transmission) for the $n$th time window and the
$k$th spatial multiplexer arm.

Now we describe, in comparison, the other case in which the logic of
the spatial multiplexer waits until any heralding photons are detected
somewhere in the system. Then it automatically routes the first
arriving heralded photons to the output. In the case when multiple
detectors click in the same time window, the time multiplexer with the
lowest labeling number will be directed to the output. The main
difference between this approach and the previous one is that the
logic does not wait until the end of the time period; at the very
first detection of heralding photons the whole system shuts. The
output probabilities to be compared with those in Eq.~\eqref{eq:ftm0}
\begin{multline}
\label{eq:ftb0}
P^{(i)}=P_0^{MN}\delta_{i,0}+\\ \sum_{j=1}^{\infty}\sum_{k=1}^{M}\sum_{n=1}^N \binom{j}{i} P_0^{M(n-1)}P_0^{(k-1)} P_j V_{nk}^i(1-V_{nk})^{j-i}
\end{multline}
Note the difference in the powers of $P_0$ compared
to the previously described priority logic. In this case, the 
$M(n-1)$st power means that no heralding events occurred in the first $n-1$
time windows in any of the sources. The $(k-1)$st power says that
the $k$th source provided a photon pair in the $n$th time window.

Equations \eqref{eq:ftm0} and \eqref{eq:ftb0} are capable of describing
any kind of combined multiplexer operating with the corresponding
priority logic. These expressions are valid even for arrangements containing
spatial multiplexers having configurations differing from the one presented
in Fig.\ \ref{f:combinedfig}, e.g., for the chained scheme presented in Ref.\ \cite{Bonneau}.
In that scheme of spatial multiplexer the number of inputs $M$ is arbitrary.
For $M=1$ or $N=1$ these equations are able to describe the standalone time
and spatial multiplexers, respectively.
The explicit form of the probabilities $P_j$ are determined by the properties
of the detector and the nonlinear source while the parameters $V_{nk}$ depend
on the practical implementation of the spatial and time multiplexers, that is,
on the parameters of the used optical elements and the geometry of the system.
Using Eqs.\ \eqref{eq:ftm0} and \eqref{eq:ftb0} one can optimize the combined
multiplexers in order to produce maximal single-photon probability.

To proceed, let us unfold all these parameters for a particular arrangement
that can be realized in bulk optics.
We will analyze such setups in detail in the next section.
We assume the use of standard threshold detectors with no photon
number resolution and detection efficiency $V_D$. The probability that
$j$ photons enter the arrangement upon an idler detection event from
the signal arm of the nonlinear photon pair source heralded by such a detector is evaluated
e.g.\ in Ref.~\cite{Adam}, and it reads
\begin{eqnarray}
{P_{0}}&=&\sum_{k=0}^{\infty}\binom{k}{0}{P^{(k)}}' V_D^0(1-V_D)^k,\nonumber\\
{P_{j}}&=&{P^{(j)}}'\sum_{k=0}^{j-1}\binom{j}{j-k}{V_D}^{j-k}\,{\left(1-V_D\right) }^{k}\,.\label{eq:Pin:det}
\end{eqnarray}
In Eq.~\eqref{eq:Pin:det}, ${P^{(k)}}'$ denotes the probability of the
generation of exactly $k$ photon pairs by a single nonlinear source within
a time window. In case of a single-mode SPDC it follows a thermal distribution
\cite{virally, silberhorn}
\begin{equation}
{P^{(k)}}'=\frac{(\lambda/N)^k}{(1+\lambda/N)^{k+1}},
\end{equation}
where $\lambda$ is the mean number of photon pairs arriving in all the
$N$ time windows. If multiple modes are accessible in the parametric
process, the probability of obtaining $k$ photons in the output field
follows Poissonian statistics instead \cite{virally, silberhorn}:
\begin{equation}
{P^{(k)}}'=\frac{(\lambda/N)^k}{k!}\exp\left(-\frac{\lambda}{N}\right).
\end{equation}
Eqs.~\eqref{eq:Pin:det} are valid in both cases.
In the following we assume SPDC sources producing photon pairs with Poissonian distribution.

Let us turn our attention to the calculation of the total transmission
probability, that is, the net transmission probability $V_{nk}$ for the $n$th time window and $k$th spatial
multiplexer arm. This quantity can be obtained in a
product form
\begin{equation}
V_{nk}=V_nV_k,\label{eq:Vnk}
\end{equation}
where $V_n$ denotes the transmission probability corresponding to the
$n$-th time window, and $V_k$ is the transmission probability of the
$k$-th spatial arm. In order to achieve the highest performance, as
inputs of the spatial multiplexer we consider those kind of bulk time
multiplexers based on binary division which have been analyzed in our previous paper~\cite{Adam}.
The transmission probability corresponding to the $n$th time window for such a setup (see Figs.~3 and 4 in Ref.~\cite{Adam}) reads
\begin{equation}
V_n=V_r^hV_t^{(l-h)}V_p^{(N-n)/N}V_b,\label{e:bulktime}
\end{equation} 
where $h$ is the Hamming weight of $N-n$ (the number of ones in its
binary representation), and $l=\log_2N$. The coefficient $V_b$ is a basic generic
transmission, independent of the $n$th time window, which
may be due to, e.g. the loss of the optical switches controlling the
path of the signal photon, etc. The reflection and transmission
efficiencies of the polarization beam splitters are denoted by $V_r$
and $V_t$, respectively. We remark that in our analysis in general we reasonably suppose that the polarizing beam splitters used in the spatial and time multiplexers are identical. Therefore we use the same notation for the transmission and reflection efficiencies $V_t$ and $V_r$ for both multiplexers. In certain cases when we assume in our analysis that these parameters differ for the spatial multiplexer we use the notation $V_{t,S}$ and $V_{r,S}$ in our calculations for the parameters of the PBS used in the spatial multiplexer.

The signal photons may be absorbed or
scattered out during the propagation in the medium. This loss is taken
into account with the propagational transmission efficiency $V_p$. The value of
$V_p$ corresponds to the longest delay which can be introduced by the
bulk time multiplexer.

Recall that the spatial multiplexer under consideration is built up
from photon routers depicted in Fig.~\ref{f:router}. We consider each
router to be identical. When building up the multiplexer from the
routers according to the scheme in Fig.~\ref{f:combinedfig}, the role
assignment of the inputs of each router (that is, which one is
considered as input 1 and which one as input 2) may depend on the
actual experimental scenario. Therefore the transmission
characteristics of a given arm shall depend on the particular setup,
but
the transmission probability $V_k'$ originated from the reflection and transmission connected to the PBS will always be described by a product of the form $V_r^qV_t^s$,
where $q$ is the total number of reflections, and $s$ is that of the
transmissions in the given arm.
Moreover, as in the case of $M$ spatial arms we always have
$m=\log_2 M$ ``levels'' of the system, and $q+s=m$, the final set of
possible transmissions is the same, but it arises in an
order depending on the above mentioned particular choice. 
In order to configure priority logics we evaluate all these data and put them into a descending order.
Then we relabel the arms according to this new ordering.
Assuming that $V_t < V_r$, the transmission probability $V_k'$ for the $k$th arm (according to the new ordering) in the spatial multiplexer is given by
\begin{equation}
\begin{alignedat}{3}
& V_k' = V_r^m, \quad && \text{if} \quad && k=1 \\ 
& V_k' = V_r^{m-1}V_t, \quad &&  \text{if} \quad &&  \binom{m}{0}<k\leq \binom{m}{0}+\binom{m}{1}  \\ 
& \vdots \quad && \vdots \quad && \vdots  \\ 
&V_k' = V_t^m \quad && \text{if} \quad && \sum_{i=0}^{m-1} \binom{m}{i}<k\leq \sum_{i=0}^{m}\binom{m}{i} \\ 
\end{alignedat}
\end{equation}
Note that for several values of $k$ the transmission probability $V_k'$ can be the same. Basically the
binomial coefficients of $(V_r + V_t)^m$ gives us how many times a
specific combination of loss of the form $V_r^qV_t^s$ appears for
a given $m$.

Another loss to be taken into account is due to the propagation
through the medium of the spatial multiplexer. We describe it with a propagational
transmission efficiency $V_{p,S}$. It depends on the size of the
combined system. Let $V_{p0,S}$ stand for the default transmission
efficiency corresponding to one level in the spatial
multiplexer. Thus the propagational transmission efficiency can be written as
\begin{equation}
V_{p,S}=V_{p0,S}^{\log_2(M)}.\label{e:spat_trans}
\end{equation}
The transmission probability corresponding to the $k$th arm of the spatial multiplexer is
\begin{equation}
V_k=V_{p,S}V_k',\label{eq:Vk}
\end{equation}
that is, the product of the two discussed quantities. Equations \eqref{eq:Vnk}-\eqref{eq:Vk} define explicitly the value of the net transmission probabilities $V_{n,k}$ in Eqs.\ \eqref{eq:ftm0} and \eqref{eq:ftb0}.

\section{Optimal combined multiplexers}\label{s:res}

Here we present our results regarding the optimization of the bulk
optical combined multiplexer described in the previous section. Within the
described framework, the optimization of a combined multiplexer
consists in the following. We fix a set of loss parameters that
describes the system. There are three parameters remaining which can
be considered as variables of the optimization procedure: the input
mean photon number $\lambda$, the number of time multiplexers $M$ and
the number of multiplexed time windows $N$. The next step is to find
$\lambda_{\text{opt}}$ for each combination of $M$ and $N$, for which the
single-photon probability is the highest. The absolute maximum of
these probabilities can be found by choosing $M$ and $N$ that
maximize it. The reason behind the existence of this optimum is that
while the increasing system size improves the efficiency of
multiplexing in principle, but the role of the losses increases
simultaneously, deteriorating this improvement.

In order to determine the maximal single-photon probability that can be realized by the considered combined multiplexers, first we consider the values of the loss parameters available in state-of-the-art experiments using bulk optical elements.
For polarization beam splitters $V_r = 0.996$ reflection and $V_t = 0.97$ transmission
efficiencies are generally feasible \cite{pbs2}. 
In Ref.~\cite{pbs1} an ultracompact high-efficiency polarization beam splitter was proposed with $V_t=0.99$. It is likely that this device with such a high transmission efficiency will be realized soon.
The transmission efficiency describing the loss due to
the propagation in the whole medium of the time multiplexer can be taken to be
$V_p=0.95$ according to Ref.~\cite{Adam}, but a bit higher values seem to be realizable as well.
The loss due to propagation in the spatial multiplexer depends
strongly on actual experimental realization of the given multiplexer,
thus it is not possible to give a generally accurate estimate.
We consider the value of the corresponding transmission efficiency assigned to one router unit to be $0.985\leq V_{p0,S}\leq 0.995$. In the following calculations we suppose without loss of
generality the value of the basic transmission efficiency $V_b=1$, and we
consider threshold detectors with an efficiency of $V_D=0.9$.
\begin{table}[!tb]
\caption{\label{tab:sep} Maximal single-photon probabilities $P_{T,\max}^{(1)}$ and $P_{S,\max}^{(1)}$ of standalone time and spatial multiplexers, and the number of multiplexed time windows $N_{T,\text{opt}}$ and spatially multiplexed SPDC sources $M_{S,\text{opt}}$ at which they can be achieved.}
\begin{center}
\begin{ruledtabular}
\begin{tabular}{lcccc|cc|cr}
No. & $V_r$ & $V_t$ & $V_p$ & $V_{p0,S}$ & $P_{T,\max}^{(1)}$ & $N_{T,\text{opt}}$ & $P_{S,\max}^{(1)}$ & $M_{S,\text{opt}}$ \\\hline
1. & 0.990 & 0.97 & 0.95 & 0.985 & 0.832 & 128 & 0.800 & 64\\
2. & 0.990 & 0.97 & 0.97 & 0.990 & 0.846 & 128 & 0.822 & 64 \\
3. & 0.993 & 0.97 & 0.96 & 0.985 & 0.850 & 128 & 0.809 & 64\\
4. & 0.996 & 0.97 & 0.95 & 0.990 & 0.854 & 128 & 0.842 & 128\\
5. & 0.996 & 0.98 & 0.95 & 0.990 & 0.874 & 128 & 0.857 & 128\\
6. & 0.996 & 0.99 & 0.95 & 0.990 & 0.899 & 256 & 0.873 & 128\\
7. & 0.996 & 0.99 & 0.96 & 0.995 & 0.907 & 256 & 0.904 & 256
\end{tabular}
\end{ruledtabular}
\end{center}
\end{table}
In Tab.~\ref{tab:sep} we have listed the maximal single-photon probabilities of bulk optical time and spatial multiplexers optimized separately using the described range of loss parameters that can be considered as experimentally feasible.
It appears that a single-photon probability as high as 80-90\% can be achieved.
The table also shows that for
higher maximal single-photon probabilities of standalone time and spatial multiplexers $P_{T,\max}^{(1)}$ and $P_{S,\max}^{(1)}$
the number of multiplexed time windows $N_{\text{opt}}$ and spatially multiplexed SPDC sources $M_{\text{opt}}$ at which these maximums can be achieved are also higher.
As we already mentioned in the introduction, the growing
number of required SPDC sources for the optimal performance appears as a drawback in an experimental
implementation, while the increase of the time windows introduces a
limitation in the achievable repetition rate.

Now we turn our attention to combined multiplexing. In order to reveal
the general characteristics of the system, it is worth to distinguish three
cases determined by the relation between the maximal
single-photon probability of the spatial multiplexer
$P_{S,\max}^{(1)}$ and that of the time multiplexer
$P_{T,\max}^{(1)}$.  Either of them may outperform the other, or the
single-photon probabilities can be roughly equal.

\begin{figure}[!tb]
\includegraphics[width=\columnwidth]{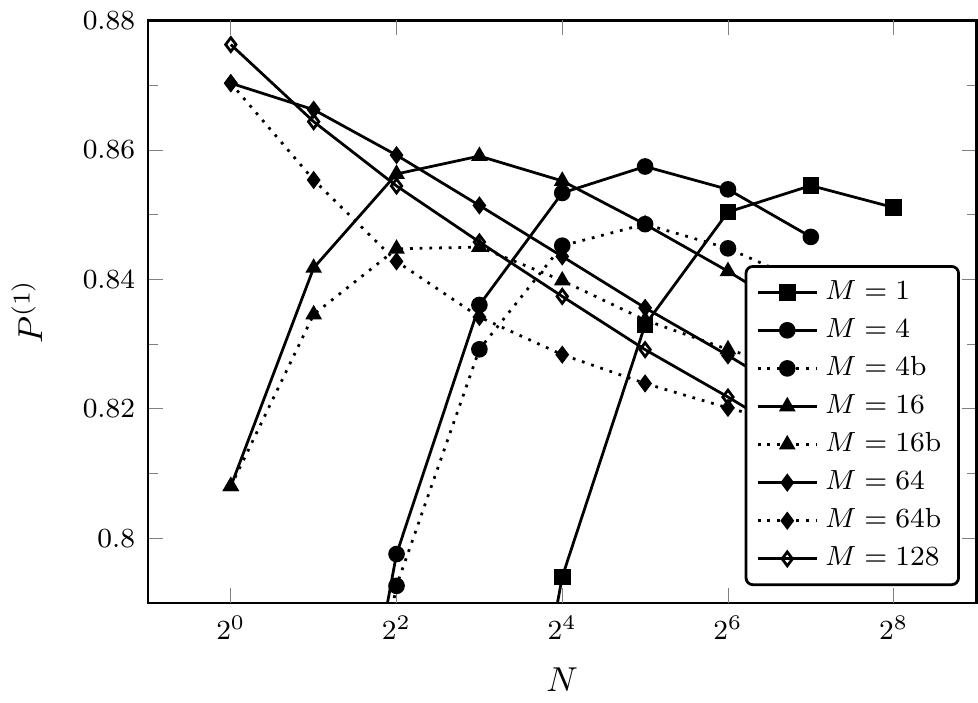}
\caption{\label{f:spath} The achievable maximal single-photon probability $P^{(1)}$ at the optimal choice of the input mean photon number $\lambda_{\text{opt}}$ as a function of the number of multiplexed time windows $N$ on semi-logarithmic scale for various number of spatially multiplexed time multiplexers $M$. Loss parameters of the analyzed multiplexer are the following: $V_r=0.996$, $V_t=0.97$, $V_p=0.95$, $V_{p0,S}=0.996$, $V_D=0.9$, $V_b=1$. Points connected with continuous lines correspond to single-photon probabilities of combined multiplexers operating with the priority logic choosing the photon in the arm of the spatial multiplexer with the lowest loss and points with dotted lines to the one which simply routes the first arriving photon to the output.}
\end{figure}

\begin{figure}[!tb]
\includegraphics[width=\columnwidth]{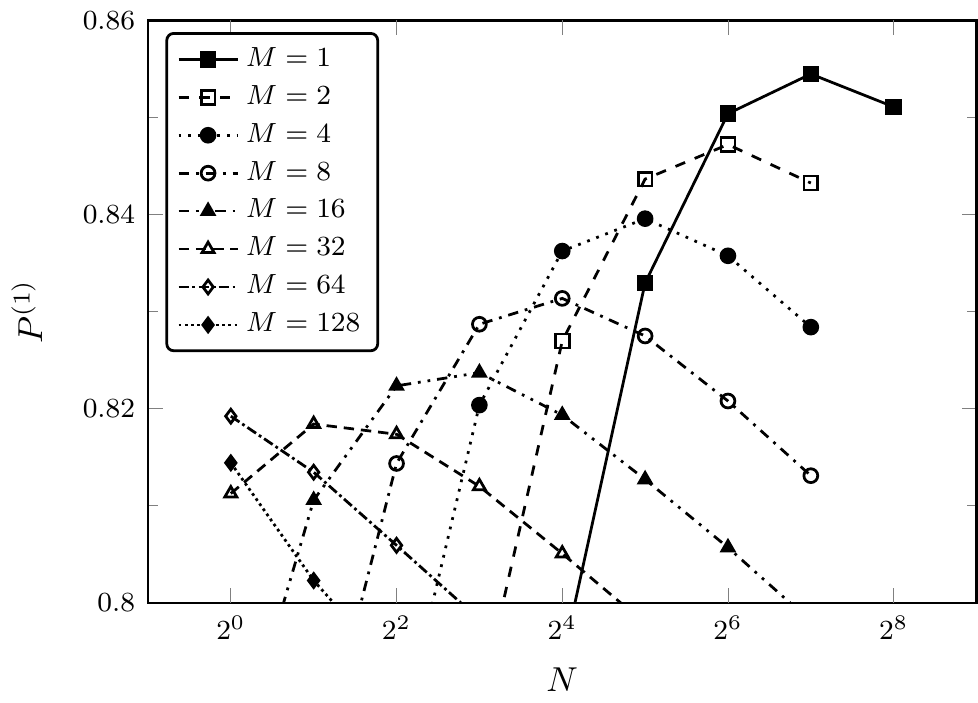}
\caption{\label{f:timh} The achievable maximal single-photon probability $P^{(1)}$ at the optimal choice of the input mean photon number $\lambda_{\text{opt}}$ as a function of the number of multiplexed time windows $N$ on semi-logarithmic scale for various number of spatially multiplexed time multiplexers $M$. Loss parameters of the analyzed multiplexer are the following: $V_r=0.996$, $V_t=0.97$, $V_p=0.95$, $V_{p0,S}=0.985$, $V_D=0.9$, $V_b=1$.}
\end{figure}

\begin{table*}[!tb]
\caption{\label{tab:effect1}
Maximal single-photon probabilities $P_{C,\max}^{(1)}$ of the combined multiplexers and the number of multiplexed time windows $N_{C,\text{opt}}$ and spatially multiplexed SPDC sources $M_{C,\text{opt}}$ at which they can be achieved for various loss parameter combinations.
The maximal single photon probabilities $P_{T,\max}^{(1)} = P_{S,\max}^{(1)} = P_{T,S,\max}^{(1)}$ of the spatial and time multiplexers if optimized themselves are also presented.
The first three rows show cases  where combined multiplexing does not enhance the maximal single-photon probabilities of spatial and time multiplexers.
The second and the last three rows present cases where combined multiplexing leads to a slightly higher and a definitely higher maximal single-photon probability, respectively.
}
\begin{ruledtabular}
\begin{tabular}{lcccccc|cccccr}
No. & $V_t$ & $V_r$ & $V_p$ & $V_{t,S}$ & $V_{r,S}$ & $V_{p0,S}$ & $P_{T,S,\max}^{(1)}$ & $M_{S,\rm opt}$& $N_{T,\rm opt}$&$P_{C,\max}^{(1)}$ & $M_{C,\rm opt}$ & $N_{C,\rm opt}$ \\\hline
1. & 0.970 & 0.996 & 0.9500 & 0.970 & 0.996 & 0.9922 & 0.8545 & 128 & 128 & 0.8531 & 2 & 64 \\
2. & 0.988 & 0.991 & 0.9589 & 0.988 & 0.991 & 0.9950 & 0.8784 & 128 & 256 & 0.8784 & 2 & 128\\
3. & 0.988 & 0.992 & 0.9568 & 0.990 & 0.991 & 0.9949 & 0.8812 & 128 & 256 & 0.8806 & 2 & 128\\\hline
4. & 0.988 & 0.990 & 0.9507 & 0.988 & 0.990 & 0.9940 & 0.8683 & 128 & 128 & 0.8684 & 2 & 64\\
5. & 0.990 & 0.996 & 0.9297 & 0.986 & 0.993 & 0.9950 & 0.8834 & 128 & 256 & 0.8840 & 2 & 128\\
6. & 0.990 & 0.996 & 0.9508 & 0.990 & 0.996 & 0.9943 & 0.8996 & 256 & 256 & 0.8999 & 2 & 128\\\hline
7. & 0.970 & 0.993 & 0.9606 & 0.980 & 0.993 & 0.9910 & 0.8506 & 128 & 128 & 0.8475 & 2 & 64\\
8. & 0.980 & 0.993 & 0.9656 & 0.990 & 0.996 & 0.9901 & 0.8740 & 128 & 128 & 0.8720 & 2 & 64\\
9. & 0.980 & 0.996 & 0.9655 & 0.990 & 0.992 & 0.9950 & 0.8860 & 128 & 256 & 0.8822 & 2 & 128\\\hline
10. & 0.980 & 0.990 & 0.9501 & 0.970 & 0.996 & 0.9917 & 0.8516 & 128 & 128 & 0.8541 & 2 & 64\\
11. & 0.990 & 0.991 & 0.9493 & 0.980 & 0.995 & 0.9940 & 0.8762 & 128 & 256 & 0.8799 & 4 & 64\\
12. & 0.990 & 0.993 & 0.9518 & 0.980 & 0.996 & 0.9951 & 0.8869 & 128 & 256 & 0.8906 & 4 & 64
\end{tabular}
\end{ruledtabular}
\end{table*}

\begin{figure}[!tb]
\includegraphics[width=\columnwidth]{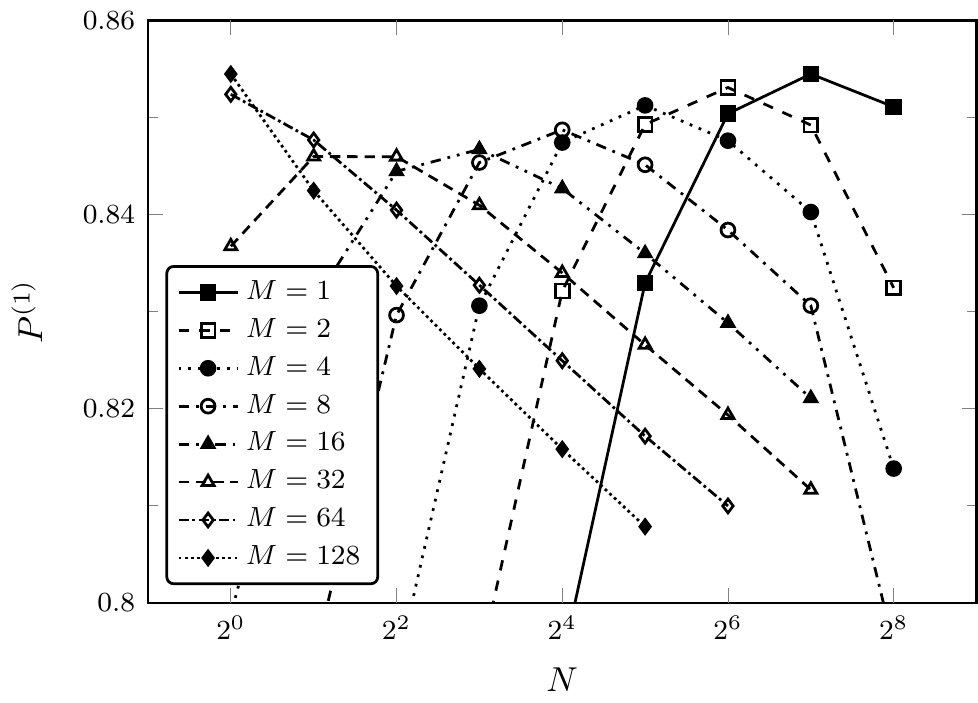}
\caption{\label{f:same1} The achievable maximal single-photon probability $P^{(1)}$ at the optimal choice of the input mean photon number $\lambda_{\text{opt}}$ as a function of the number of multiplexed time windows $N$ on semi-logarithmic scale for various number of spatially multiplexed time multiplexers $M$. Loss parameters of the analyzed multiplexer are the following: $V_r=V_{r,S}=0.996$, $V_t=V_{t,S}=0.97$, $V_p=0.95$, $V_{p0,S}=0.9922$, $V_D=0.9$, $V_b=1$.}
\end{figure}

\begin{figure}[tb]
\includegraphics[width=\columnwidth]{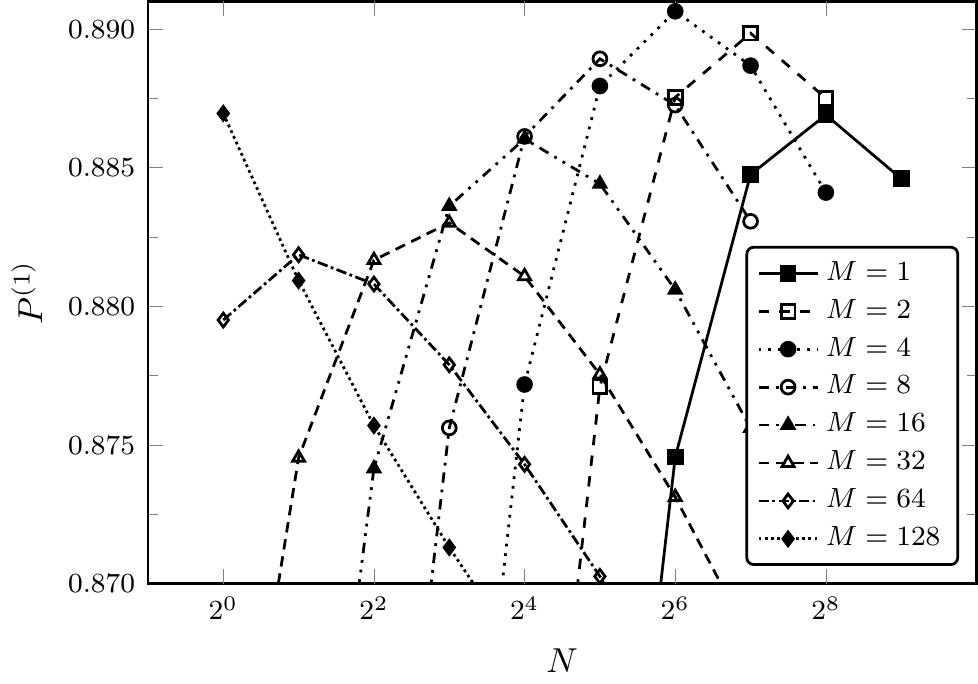}
\caption{\label{fig:no9} The achievable maximal single-photon probability $P^{(1)}$ at the optimal choice of the input mean photon number $\lambda_{\text{opt}}$ as a function of the number of multiplexed time windows $N$ on semi-logarithmic scale for various number of spatially multiplexed time multiplexers $M$. Loss parameters of the analyzed multiplexer are the following: $V_r=0.993$, $V_{r,S}=0.996$, $V_t=0.99$, $V_{t,S}=0.98$, $V_p=0.95$, $V_{p0,S}=0.995$, $V_D=0.9$, $V_b=1$.}
\end{figure}

Let us first analyze the case when the maximal single-photon probability of the spatial multiplexer is higher than that of the time multiplexer, that is, $P_{S,\max}^{(1)} > P_{T,\max}^{(1)}$.
In Fig.~\ref{f:spath} we show the achievable maximal single-photon probability $P^{(1)}$ at the optimal choice of the input mean photon number $\lambda_{\text{opt}}$ as a function of the number of multiplexed time windows $N$ for various number of spatially multiplexed time multiplexers $M$ for an experimentally feasible set of loss parameters.
The high performance $P_{S,\max}^{(1)}=0.8763$ of the spatial multiplexer is ensured by choosing a transmission efficiency as high as $V_{p0,S}=0.996$.
In this figure the curve $M=1$ corresponds to standalone time multiplexers while the points at $N=2^0=1$ are calculated for standalone spatial multiplexers.
The figure shows results for both of the considered priority logics treated in Sec.~\ref{s:stat}.
The points connected with dotted lines correspond to the logic which routes the signal photons from the first detected heralding event to the output.
The points connected with continuous lines correspond to the improved logic choosing the spatial arm of the lowest loss.
For $M=1$ (no spatial multiplexing) or $N=1$ (no time multiplexing) obviously the two logics produce the same performance. 
For the other choices of $N$ and $M$, combined multiplexers operating with the priority logic choosing the photon in the arm of the spatial multiplexer with the lowest loss produce always higher single-photon probabilities.
Let us note here that we have made this comparison for all the following calculations, and we have found that the improved logic always outperforms the simpler one.
Therefore, while we emphasize this fact here, we omit the details of this comparison in what follows.

The absolute maximal single-photon probability in Fig.~\ref{f:spath} is at $M=128$ and $N=1$.
This suggests that the best choice would be not to apply time multiplexing at all.
However, the corresponding spatial multiplexer would involve 128 SPDC sources in the considered bulk optics setup, which is clearly unreasonable in practice.
Combined multiplexing, on the other hand, can solve the issue of system size: single-photon probabilities over 86\% can be achieved, for instance, with just 4 SPDC sources.
Thus in this case, combined multiplexing enhances the achievable maximal
single-photon probability $P^{(1)}=85.4\%$ of a single bulk time
multiplexer. Notice that single-photon probabilities
over 86\% can be achieved with less than $N=128$ multiplexed time
windows. As a consequence of this decrease of the number of time
windows, higher repetition rates can be achieved with
combined multiplexers, as compared to optimized single time
multiplexers.

Moreover, the described advantage of the combined multiplexing is valid for several configurations with different number of spatially multiplexed sources $M$ and multiplexed time windows $N$. For such points the single-photon probability of the combined system is between the maximal probabilities of the standalone spatial and time multiplexers but the values of $M$ and $N$ in the combined system are smaller than the values $M_{S,\text{opt}}$ and $N_{T,\text{opt}}$ in the optimized standalone systems.
In addition, the single-photon probabilities are significantly higher than the ones
that can be achieved with the suboptimal use of the standalone spatial multiplexer when the number of multiplexed units $M$ is far below the optimized value $M_{S,\text{opt}}$, that is, $M\leq M_{S,\text{opt}}/4$.

Now let us consider the complementary case when the maximal single-photon probability of the spatial multiplexer is lower than that of the time multiplexer, that is, $P_{S,\max}^{(1)} < P_{T,\max}^{(1)}$.
In Fig.~\ref{f:timh} the achievable maximal single-photon probability $P^{(1)}$ is plotted at the optimal choice of the input mean photon number $\lambda_{\text{opt}}$ as a function of the number of multiplexed time windows $N$, for various numbers of spatially multiplexed time multiplexers $M$, for an experimentally feasible set of loss parameters. In this case the propagational transmission efficiency of the spatial multiplexer is chosen
to be $V_{p0,S}=0.985$, and it has maximal single-photon probability
$P_{S,\max}^{(1)}=82\%$ at $M=64$.
It appears that combined multiplexing
does not enhance the absolute maximum of single-photon probability
($P_{T,\max}^{(1)}=85.4\%$ at $M=1$, $N=128$) at all in this case.
On the other hand, when the number of time windows $N$ is far below the optimized value $N_{T,\text{opt}}$, that is, $N\leq N_{T,\text{opt}}/4$, there are lots of combinations of $M$ and $N$ for which the single-photon probability is higher than one can achieve by this suboptimal use of a standalone time multiplexer.
Therefore the benefit of the application of the spatial multiplexer is the possible enhancement of the repetition rate as described before, without the relevant decrease of the single-photon probabilities.

We have analyzed the two cases described above for a variety of different sets of loss efficiencies.
Without going into details we remark here that we found the described behavior for all of the choices.

Finally, let us analyze the third possibility when the maximal single-photon probabilities of the spatial and time multiplexers, provided that they are optimized themselves, are equal within a given precision, that is, $P_{S,\max}^{(1)} = P_{T,\max}^{(1)}=P_{T,S,\max}^{(1)}$.
We have performed simulations for several combinations of the loss parameters ensuring this equality.
We have found that the maximal single-photon probability $P_{C,\max}^{(1)}$ of the combined system can be slightly lower or higher than the maximal single-photon probability $P_{T,S,\max}^{(1)}$ of the spatial and time multiplexers.
The difference is generally so small that it cannot be detected in an experiment. As a consequence these quantities can be considered as roughly equal.
For certain parameter sets these probabilities are really equal at the given precision.
Beside this behavior we have found some rather special sets of loss parameters for which the single-photon probability of the combined system is definitely, yet not significantly lower or higher than that of the spatial or time multiplexed systems separately.
Table \ref{tab:effect1} shows some examples for all these behaviors.
In this table we present the maximal single-photon probabilities $P_{C,\max}^{(1)}$ of the combined multiplexers and the number of multiplexed time windows $N_{\text{opt}}$ and spatially multiplexed SPDC sources $M_{\text{opt}}$ at which they can be achieved for various loss parameter combinations.
The maximal single photon probabilities $P_{T,S,\max}^{(1)}$ of the spatial and time multiplexers if optimized themselves are also presented.

Rows 1-6 of Tab.~\ref{tab:effect1} contain cases when the optimized performance of the combined multiplexer is roughly equal to that of the standalone spatial and time multiplexers.
The first three rows show examples for parameters for which the single-photon probability of the combined multiplexer $P_{C,\max}^{(1)}$ is slightly lower, while for the parameters in the second three rows it is slightly higher than that of the standalone multiplexers $P_{T,S,\max}^{(1)}$.
We note that for the parameter set presented in row 2 all the single-photon probabilities are equal at the given precision, although the previous statement is true for this example.
Rows 7-9 of Tab.~\ref{tab:effect1} show examples for parameters for which the single-photon probability of the combined multiplexer $P_{C,\max}^{(1)}$ is definitely lower, while for the parameters in the last three rows it is definitely higher than that of the standalone multiplexers $P_{T,S,\max}^{(1)}$. The difference in the probabilities exceeds 0.002 (0.2\%). Such a behavior occurs only if at least one of the transmission and reflection efficiencies of the applied PBS-s differs for the time and spatial multiplexers.
An interesting feature that can be deduced from Tab.~\ref{tab:effect1} is that the product of the optimal number of spatially multiplexed time multiplexers $M_{C,\rm opt}$ and the optimal number of multiplexed time windows $N_{C,\rm opt}$ for the combined system is equal to the optimal number of multiplexed time windows $N_{T,\rm opt}$ for the standalone time multiplexed source, that is, $M_{C,\rm opt}N_{C,\rm opt}=N_{T,\rm opt}$.
This property is valid for other combined configurations, presented in Figs.~\ref{f:spath}-\ref{fig:no9}, for certain sets of the number of spatially multiplexed time multiplexers $M$ and the number of time windows $N$ ensuring the best performance of the given combined multiplexed system. In these figures, such values of $M$ are $M\leq 32, 16, 16$ and $32$, respectively. Table~\ref{tab:effect1} also shows, taking into account previous considerations as well, that by using combined multiplexing systems realized in bulk optics single-photon probabilities between 85\% and 89\% can be achieved experimentally.

Figs.~\ref{f:same1} and \ref{fig:no9} show the achievable maximal single-photon probability $P^{(1)}$ at the optimal choice of the input mean photon number $\lambda_{\text{opt}}$ as a function of the number of multiplexed time windows $N$ for various numbers of spatially multiplexed time multiplexers $M$ for the loss parameters presented in the first and the last rows of Tab.~\ref{tab:effect1}, respectively.
In Fig.~\ref{fig:no9} one can see that beside the point corresponding to the maximal single-photon probability ($M_{C,\text{opt}}=4$ and $N_{C,\text{opt}}=64$) there are other $(M,N)$ pairs for this configuration [(2,64), (2,128), (2,256), (4,32), (4,128), (8,32), (8,64)] at which the single-photon probability exceeds the maximal single-photon probabilities of the standalone spatial and time multiplexers.
These figures also show that there are several choices of $M$ and $N$ for which the single-photon probabilities are higher than one can achieve by suboptimal use of a standalone spatial or time multiplexer. Furthermore, these are not significantly lower than the maximal value.
The aforementioned benefits of the combined approach, namely the decrease of the required SPDC sources and improvement of
the achievable repetition rate, are also present in these cases.
Finally, as we have already mentioned before, combined multiplexing allows continuous
pumping of the system, which appears as an additional advantage of this arrangement.

\section{Conclusions}\label{s:concl}
We have studied periodic single-photon sources based on combined
multiplexing, in which the outputs of several time multiplexers are
spatially multiplexed. We have set up a general framework for the
description and optimization of such devices. Such systems can be
realized most efficiently in bulk optics. We have pointed out that due to
the asymmetry present in such a setup, it is possible to design an
improved priority logic for the spatial part of the multiplexer.

We have shown that combined multiplexing systems can be
optimized in order to achieve maximal single-photon probability for
various sets of loss parameters by the appropriate choice of the
number of spatially multiplexed time multiplexers, the number of
multiplexed time windows and the input mean photon number.

According to our results concerning bulk optical combined multiplexers,
if either the spatial or the time multiplexer
outperforms the other, the combination can achieve an improvement
compared to the worse one, even though it cannot be superior to the
absolute maximum defined by the better one.
If the spatial and time multiplexers themselves have a
similar optimum performance, their combination may yield an enhanced
single-photon probability in some special cases.

Finally, let us note that the performance of the combined multiplexers
is generally higher than that of the standalone time or spatial
multiplexers below optimized system size.  More importantly, the
combination can lead to a decrease in the number of the required SPDC
sources or a possible increase of the achievable repetition rate of
the system compared to the standalone use of the optimized spatial or
time multiplexers, while still maintaining a relatively high
single-photon probability.  All these features of combined
multiplexing can be essential from the point of view of experiments.

\begin{acknowledgments}
We thank the support of the National Research Fund of Hungary OTKA (Contract No.\ K83858) and the German-Hungarian collaboration project TKA-DAAD project No.\ 65049.
\end{acknowledgments}

\end{document}